\documentclass{article}

\usepackage{graphicx}           
\usepackage{bm}                 
\usepackage{amsmath}            
\usepackage{verbatim}           
\usepackage{amsthm}             
\theoremstyle{plain}            
\newtheorem{theorem}{Theorem}

\newtheorem{lemma}{Lemma}

\title{Three routes to the exact asymptotics for the one-dimensional
quantum walk}

\author{Hilary A. Carteret \\
Department of Combinatorics and Optimization \\
             University of Waterloo \\ 
Waterloo, Ontario, N2L 3G1 \\ 
Canada \\
email: hcartere@iqc1.math.uwaterloo.ca
\and Mourad E. H. Ismail \\
Department of Mathematics \\ 
University of Central Florida \\ 
Orlando, Florida 32816 \\ 
USA \\
email: ismail@math.usf.edu
\and Bruce Richmond \\
Department of Combinatorics and Optimization \\
             University of Waterloo \\ 
Waterloo, Ontario, N2L 3G1 \\ 
Canada \\
email: lbrichmo@hopper.math.uwaterloo.ca}

\date{March 18, 2003}

\begin{document}
\maketitle

\newpage 

\begin{abstract}
We demonstrate an alternative method for calculating the asymptotic 
behaviour of the discrete one-coin quantum walk on the infinite line, via 
the Jacobi polynomials that arise in the path integral representation.  
We calculate the asymtotics using a method that is significantly easier 
to use than the Darboux method.  It also provides a single integral 
representation for the wavefunction that works over the full range of 
positions, $n,$ including throughout the transitional range where the 
behaviour changes from oscillatory to exponential.  Previous analyses of 
this system have run into difficulties in the transitional range, because 
the approximations on which they were based break down here.  The fact 
that there are two different kinds of approach to this problem (Path 
Integral vs.  Schr\"{o}dinger wave mechanics) is ultimately a manifestation 
of the equivalence between the path-integral formulation of quantum 
mechanics and the original formulation developed in the 1920s.  We also 
discuss how and why our approach is related to the two methods that have 
already been used to analyse these systems.

\end{abstract}






\section{Introduction}

The discrete quantum walk has been discussed in several recent papers 
\cite{NayaknV,Ambainis01}.
The first authors to discuss the quantum random walk were 
Y. Aharonov, Davidovich and Zagury, in \cite{earliest} where they
described a very simple realization of the quantum random walk in quantum
optics. Some further early results were due to Meyer, in \cite{Meyer96}.  
He proved that for a discrete (unitary) quantum walk on the line to have
non-trivial behaviour, its motion must be assisted by an additional
``coin'' degree of freedom which is conventionally taken to be two
dimensional.  This spin-like degree of freedom is sometimes called the
{\emph{chirality,}} and can take the values {\small{RIGHT}} and
{\small{LEFT}}, or a superposition of these.  Meyer therefore considered
the wave function as a two component vector of amplitudes of the particle
being at point $n$ at time $t$.
Let
\begin{equation}\label{psidef}
  \Psi(n,t)=
   \begin{pmatrix}
     \psi_{L}(n,t) \\
     \psi_{R}(n,t)
   \end{pmatrix}
\end{equation}
where we shall label the chirality of the top component {\small{LEFT}} and 
the bottom {\small{RIGHT}}.  This paper is concerned with the dynamics 
of a test particle performing an unbiased quantum walk on the integer 
points on the line.  At each time step the chirality of the particle 
evolves according to a unitary Hadamard transformation 
\begin{align}\label{spinstep}
|\rm{R}\rangle \mapsto \frac{1}{\sqrt{2}}(|\rm{R}\rangle + |\rm{L}\rangle) \\
|\rm{L}\rangle \mapsto \frac{1}{\sqrt{2}}(|\rm{R}\rangle - |\rm{L}\rangle) 
\end{align}
and then the particle moves according to its (new) chirality state.  
Therefore, the particle obeys the recursion relations
\begin{align}\label{recursion}
  \Psi_{\rm{L}}(n,t+1)&=-\frac{1}{\sqrt{2}}\Psi_{\rm{L}}(n+1,t)
 +\frac{1}{\sqrt{2}}\Psi_{\rm{R}}(n-1,t) \\
  \Psi_{\rm{R}}(n,t+1)&=\frac{1}{\sqrt{2}}\Psi_{\rm{L}}(n+1,t)
+\frac{1}{\sqrt{2}}\Psi_{\rm{R}}(n-1,t).
\end{align} 
Meyer and subsequent authors have considered two approaches to the 
Hadamard walk, the {\emph{Path Integral}} and {\emph{Schr\"{o}dinger}} 
approaches, which reflect two complementary ways of formulating quantum 
mechanics \cite{FeynmanHibbs}.  We refer to the paper by Ambainis, Bach, 
Nayak, Vishwanath and Watrous \cite{Ambainis01} for proper definitions and 
references.  We shall refine the asymptotic analysis of this paper. 

\begin{figure}[floatfix]
  \begin{minipage}{\columnwidth}
   \begin{center}
    \resizebox{0.8\columnwidth}{!}{\includegraphics{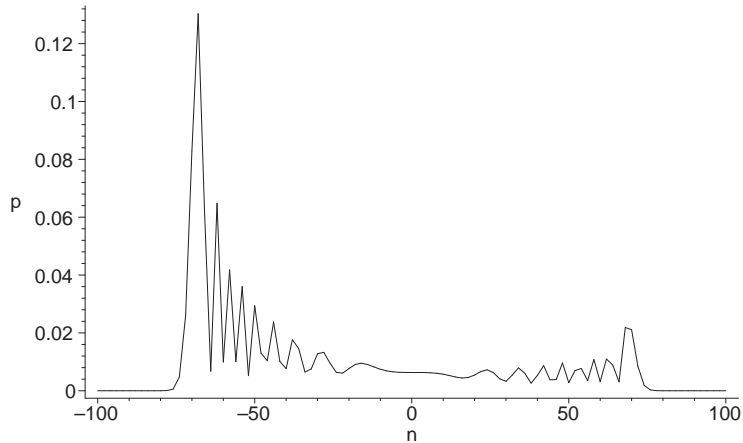}}
   \end{center}
  \end{minipage}
  \caption{The discrete quantum walk on the line.  The probability 
           distribution is shown for a walk that started at the origin
           with its coin in the state $|\rm{R}\rangle$ after it has evolved 
           for 100 steps.  The distribution is oscillatory between the two 
           peaks and decays exponentially outside that range.  The peaks 
           move away from the origin with speed $t/\sqrt{2},$ and the width 
           of the peaks also decreases with time.  By contrast, the classical 
           random walk has a Gaussian distribution, which spreads with 
           velocity $\sim \sqrt{t}.$} \label{walk}
\end{figure}

The behaviour of the Hadamard walk is very different from the classical 
random walk on the integer points on the real line.  One way of understanding 
this is as a result of quantum interference.  Destructive interference 
suppresses the probability amplitude in the vicinity of the origin, and 
constructive interference reinforces it away from the origin.  The net effect 
of this is that the quantum walk spreads out much faster. Figure \ref{walk} 
shows the discrete quantum walk on the infinite line at $t=100.$  We have only 
plotted the distribution for even values of $n$ in Figure \ref{walk}.  If a 
walk's initial distribution has its support confined to a set of nodes which 
all have the same parity (all $n$s either even or odd) the support of the 
distribution will ``tick'' between different parities at each step.

The probability distribution for the quantum walk depends not only on the 
evolution law in \eqref{spinstep}, but also on the initial conditions. 
Throughout this paper we will only consider walks that start at $n=0$ with the
coin in the initial state $|\rm{R}\rangle.$  The convention that the walk 
starts in the position $n=0$ is inherited from the motivation for studying 
these systems as toy models of quantum algorithms: the computer is always 
started with its registers in the state $|00\ldots\rangle.$  
The choice of initial coin state was made because the Hadamard walk is an 
unbiased walk \cite{NayaknV}.  This means that even though some starting 
conditions result in an asymmetric probability distribution, we can always 
find some other starting state that will produce a walk with the opposite 
bias.   
The $|\rm{R}\rangle$ starting state produces a distribution that is maximally 
biased to the right.  Likewise, if we had chosen to start the walk in the 
state $|\rm{L}\rangle,$ this would have produced a distribution that was 
maximally biased to the left.  This distribution is the exact mirror-image 
of that produced by starting in the state $|\rm{R}\rangle.$ Thus, reversing 
the starting condition just relabels $n$ to $-n.$  As the coin space is 
two-dimensional, we can now invoke the linearity of quantum mechanics, and 
note that we can obtain the behaviour for any initial condition 
$|\Phi\rangle=a|\rm{R}\rangle+b|\rm{L}\rangle$ by forming the corresponding 
linear combination of the evolutions for the initial condition basis states 
$|\rm{R}\rangle,|\rm{L}\rangle.$

\section{Related Work}

We now very briefly describe the methods previous authors have used. They
have so far followed two approaches to determine the limiting behaviour of
the $\psi$-functions as $t \rightarrow \infty$. The translational invariance 
of this problem means that it has a simple description in momentum space, 
and the Schr\"{o}dinger approach relies on that fact.  We will describe 
this in more detail in the section below.  Beginning with the recursion 
relation \eqref{recursion}, Nayak and Vishwanath \cite{NayaknV} showed
that
\begin{equation}
\tilde{\Psi}(k,t)=\left(M_{k}\right)^{t} \tilde{\Psi}(k,0)
\end{equation}
where
\begin{equation}
\tilde{\Psi}(k,0)=\sum_{n}\Psi(n,0)e^{ikn}
\end{equation}
and $M_{k}$ is the matrix
\begin{equation}
M_{k}=\frac{1}{\sqrt{2}}
\begin{pmatrix}
  -e^{-ik} & e^{-ik} \\
  e^{ik} & e^{ik}
\end{pmatrix}.
\end{equation}

They diagonalize the matrix $M_{k}$, finding the eigenvalues
$\lambda_{k}^{1}=e^{i\omega_{k}}$ and
$\lambda_{k}^{2}=e^{i(\pi-\omega_{k})}$ where
$\omega_{k}=\arcsin\left(\frac{\sin k}{\sqrt{2}}\right), \quad
\omega_{k}\in [-\pi/2,\pi/2]$. They then write the $\psi$-functions
in terms of the eigenvalues $\left(\lambda_{k}^{1}\right)^{t}$
and $\left(\lambda_{k}^{2}\right)^{t}$ and their associated eigenvectors
and formally invert the original fourier transform to obtain the closed
form integral representations for the wavefunction, 
\begin{align}
  \psi_{\rm{L}}(n,t) &= \frac{1+(-1)^{n+t}}{2}\int_{-\pi}^{\pi} 
   \frac{dk}{2\pi}
   (1+\frac{\cos k}{\sqrt{1+\cos^2k}})e^{-i(\omega_kt+kn)} \\ 
  \psi_{\rm{R}}(n,t) &= \frac{1+(-1)^{n+t}}{2}\int_{-\pi}^{\pi}
   \frac{dk}{2\pi}
   \frac{e^{ik}}{\sqrt{1+\cos^2k}}e^{-i(\omega_kt+kn)}.
\end{align}
They then approximate these, using a combination of the method of 
stationary phase in one range, and integration by parts in the 
other.  (Note that the Left-Right labelling convention in \cite{NayaknV} is 
the opposite to ours.)

\bigskip

There is another approach based on the {\emph{Path Integral}} formulation 
of quantum mechanics, that Ambainis {\emph{et al.}} discuss. The 
$\psi$-functions are expressed in terms of Jacobi polynomials (as in Lemma 
\ref{second} below) of the form
\begin{equation}
J_{j}^{(aj+\gamma,bj+\beta)}(0), \; j=(t-n)/2-1=(1-\gamma)t/2 -1.
\end{equation}
One may then derive the asymptotic behaviour of the $\psi$-functions as
$t \rightarrow \infty$ by determining the asymptotic behaviour of these
Jacobi polynomials as $m \rightarrow \infty$. This has been done in two
ways.  Ambainis {\emph{et al.}} use the approach due to Chen and Ismail 
\cite{ChenIsmail}, which employs the {\emph{Darboux}} method.  Here one 
begins with the Srivastava-Singhal generating
function
\begin{equation}
 \sum_{j=0}^{\infty} J_{j}^{(\gamma + aj,\beta + bj)}(w)z^{j}
   =(1+{\bf{u}})^{-\gamma}(1+{\bf{v}})^{-\beta}[1+(1+a){\bf{u}}
   +(1+b){\bf{v}}]^{-1}
\end{equation}
where ${\bf{u}}$ and ${\bf{v}}$ are defined to be power series in $w$ that
satisfy the equations
\begin{align}
{\bf{u}}&=-\frac{1}{2}(w+1)z(1+{\bf{u}})^{-\lambda}(1+{\bf{v}})^{-\mu-1} \\
{\bf{v}}&=-\frac{1}{2}(w-1)z(1+{\bf{u}})^{-\lambda-1}(1+{\bf{v}})^{-\mu}.
\end{align}
Chen and Ismail use the Darboux method to calculate the asymptotics of 
these Jacobi polynomials. This method starts from the idea that if
$f(z)=\sum_{n \ge 0}a_{n}z^{n}$ then the radius of convergence is 
$\mathcal{R}=\lim_{n \rightarrow \infty}
\left(\frac{1}{|a_{n}|}\right)^{1/n}$ when this limit exists.  Suppose
there is a {\emph{comparison function}} $g(z)=\sum_{n \ge 0}b_{n}z^{n}$
such that $g(z)-f(z)$ has a larger radius of convergence than $f(z),$ then
$b_{n}-a_{n}=O(s^{-n})$ where $s > \mathcal{R}.$  If the asymptotic
behaviour of $b_{n}$ is known, then since $a_{n}\sim b_{n}$ then we know
the asymptotic behaviour of $a_{n}.$  Chen and Ismail use the
Srivastava-Singhal description of the generating function for Jacobi
polynomials to determine its singularities on the radius of convergence
and give comparison functions at each singularity to determine the
asymptotic behaviour of the Jacobi polynomials. 

It is interesting to note that the reciprocals of these two singularities
(when normalized by dividing by $\left(\sqrt{2}\right)^{t}$) are the
eigenvalues that arise in the Schr\"{o}dinger method.

\section{Overview and Results}

The other way to obtain the asymptotic behaviour of these Jacobi 
polynomials is to follow the method in Gawronkski and Shawyer's paper, 
\cite{Gawyer}.  This is the third way to analyse these systems, and will 
be the way that we follow in much of this paper.  It is a refinement of 
the methods in the paper by Saff and Varga \cite{Saff}.  This uses the 
method of {\bf{steepest-descents}}.  We will outline this method below but 
for further details we recommend \cite{Arfken,Wong} and also the book by 
Olver \cite{Olver} which describes both the steepest-descent method and the 
method of Darboux very clearly.

The saddle-points that feature in this method are also related
to the eigenvalues that arise in the Schr\"{o}dinger method.  We will
detail this relationship below.   If $\zeta$ is a saddlepoint then
$\exp(h(\zeta))$ is an eigenvalue from the Schr\"{o}dinger method, for a
function $h(\zeta)$ that we define below.

\bigskip

We now describe our results.  We shall see that it is possible to obtain
explicit asymptotic expansions that are uniformly convergent.  The system 
displays two types of behaviour, with the transitions between the 
different behaviours governed by a parameter, $\alpha =n/t$. The behaviour 
changes qualitatively over three ranges, which are respectively 
$0 \le |\alpha| \le \frac{1}{\sqrt{2}}-\varepsilon,
\frac{1}{\sqrt{2}}-\varepsilon <|\alpha| <\frac{1}{\sqrt{2}}+\varepsilon,$
and $\frac{1}{\sqrt{2}} +\varepsilon \le |\alpha|<1-\varepsilon$, where
$\varepsilon$ is a positive number. Our methods give error terms of the 
form $O(t^{-N-1/2})$ where $N$ is some positive integer, and the expansion 
for each range holds uniformly. We stop at $O(t^{-5/2})$ since we have no
application for the more precise estimates. 

The first range is arguably the most interesting. Here the
asymptotic behaviour of the $\psi$-functions is oscillatory as
$t\rightarrow \infty$. We will use a result by Gawronkski and Shawyer
\cite{Gawyer} who obtained the leading term and the factor $1+O(t^{-1})$
for the error term for Jacobi polynomials.  When this result is applied to
the $\psi$-functions, we obtain a refinement of Theorem 2 of 
\cite{Ambainis01} who found the leading term.

The second, boundary, range is treated using the method of coalescing
saddle-points as described in R. Wong's book \cite{Wong}.  A uniform
asymptotic expansion is possible which involves the Airy function.  This 
is also interesting because the $\psi$-functions change from an oscillating, 
polynomially bounded asymptotic behaviour to an exponentially small
behaviour as $|\alpha|$ changes from below $1/\sqrt{2}$ to above
$1/\sqrt{2}$.  The calculation of the asymptotics for these polynomials is 
novel to the best of the authors' knowledge.  The third range is the 
immediate vicinity of $|\alpha| = 1/\sqrt{2}.$  As the behaviour of our 
integral representation is well understood, we can obtain uniformly 
convergent asymptotics thoughout this transitional range. 

\bigskip

The main interest of our results lies not in the more accurate asymptotic
expansions for the $\psi$-functions in the first and third ranges of
$\alpha,$ although Ambainis {\emph{et al.}} did ask for a uniform method 
to do the asymptotics for these ranges.  In fact, Gawronkski and Shawyer 
have already provided this method \cite{Gawyer}.  The fact that the uniform 
asymptotic behaviour for the Jacobi polynomials for the transistion from 
polynomially bounded oscillating behaviour to exponential decay can be 
found in terms of the Airy function seems rather more interesting.  It is 
somewhat surprising that this has not (to our knowledge) been published 
before for this family of polynomials: Wong's book \cite{Wong} shows how 
to obtain the asymptotic behaviour in terms of Airy functions.  Perhaps the 
reason for this omission is simply the previous lack of an application, 
which the quantum walk now provides.

\section{The Feynman Path Integral}

We begin with the Feynman path integral appproach following Meyer. We will
represent the components of the vector-valued wavefunction as a normalized
sum over signed paths.  Meyer \cite{Meyer96} proved
\begin{lemma}[Meyer \cite{Meyer96}]\label{start}
Let $-n\le t<n$.  Define $l=\frac{t-n}{2}$.  The amplitudes of position $n$ 
after $t$ steps of the Hadamard walk are:
\begin{align}
  \psi_{R}(n,t)&=\frac{1}{\sqrt{2^{t}}}
    \sum_{s} \binom{l-1}{s-1} \binom{t-l}{s}(-1)^{t-s} \\
  \psi_{L}(n,t)&=\frac{1}{\sqrt{2^{t}}}
    \sum_{s}\binom{l-1}{s} \binom{t-l}{s}(-1)^{t-s-1},
\end{align}
except for the endpoints where $t=n,$ which have to be handled separately. 
\end{lemma}
(See \cite{Meyer96} for the endpoints.  For a derivation of these 
formulae, see Appendix A of \cite{Multicoin}.)

We will follow the approach used by Ambainis {\emph{et al.}}
\cite{Ambainis01} and Meyer \cite{Meyer96} but in greater detail to
obtain the following lemma.  The standard notation in \cite{AnStegun} for
Jacobi polynomials is to write them as $J_{n}^{(\alpha,\beta)}(z),$ but to
avoid confusion with $\alpha=n/t$ we will write these as 
$J_{r}^{(u,v)}(z).$  (Note that in the following lemma, we have reintroduced 
the external phase that was omitted in \cite{Meyer96} and \cite{Ambainis01}.) 

\begin{lemma}[Ambainis {\emph{et al.}}\cite{Ambainis01}]\label{second}
When $ n \equiv t$ mod 2 and $J_{r}^{(u,v)}(z)$ denotes a Jacobi 
polynomial,
\begin{equation}
\psi_{R}(n,t)(-1)^{(t-n)/2} =
\begin{cases}
    -(\frac{t+n}{t-n})2^{-n/2-1}J_{(t-n)/2-1}^{(1,n)}(0),
     & \text{ when }0\le n<t \\
    -2^{n/2-1}J_{(t+n)/2-1}^{(1,-n)}(0), & \text{ when }-t\le n<0.
\end{cases}
\end{equation}
Also,
\begin{equation}
\psi_{L}(n,t)(-1)^{(t-n)/2} =
\begin{cases}
(-1)^{n+1}2^{-n/2-1}J_{(t-n)/2-1}^{(0,n+1)}(0), & \text{ when }0 \le n <t \\
(-1)^{n+1}2^{n/2} J_{(t+n)/2}^{(0,-n-1)}(0), & \text{ when }-t \le n < 0.
\end{cases}
\end{equation}
\end{lemma}

{\bf{Remark:}} As Ambainis {\emph{et al.}} \cite{Ambainis01} point out
\begin{align}
&\psi_{L}(-n,t)=-\psi_{L}(n-2,t),
&\psi_{R}(-n,t)=\frac{t-n}{t+n}\psi_{R}(n,t).
\end{align}

Proof of Lemma~\ref{second}: We use two formulae from Abromowitz and 
Stegun \cite{AnStegun} to prove these results. The first is the 
Pfaff-Kummer transformation \cite{AskeyRoy} (15.3.4 of \cite{AnStegun})
\begin{equation}\label{hypergeomone}
 \;_{2}F_{1}(a,b;c;z)
   =(1-z)^{-a}\;_{2}F_{1}(a,c-b;c;\frac{z}{z-1}).
\end{equation}
The second is the representation of a Jacobi polynomial as a $\;_2F_1$, 
see 15.4.6 of \cite{AnStegun} 
\begin{equation}\label{hypergeomtwo}
 \;_{2}F_{1}(-j,u+1+v+j;u+1;z) =
 \frac{j!}{(u+1)_{j}} J_{j}^{(u,v)}(1-2z).
\end{equation}
Now the first sum, say $S,$ in Lemma~\ref{start} is
$\;_{2}F_{1}(1-l,-(t-l);1;-1)$ so by \eqref{hypergeomone}
\begin{equation}
  S=2^{(t-n)/2-1} \;_{2}F_{1}(1-(t-n)/2,-1+(t-n)/2+n+1+1;1;1/2).
\end{equation}
Now by \eqref{hypergeomtwo} we obtain
\begin{equation}
S=2^{(t-n)/2-1}J_{(t-n)/2-1}^{(0,n+1)}(0).
\end{equation}
This proves the first part of Lemma~\ref{second} for $n\ge 0.$ To derive 
the second part, for $n < 0$ we see
\begin{multline}
S = \;_{2}F_{1}(-(t+n)/2,1-1+(n-t)/2;1;-1) \\ 
  = 2^{(t+n)/2} \;_{2}F_{1}(-(t+n)/2,1+(n+t)/2-n-1;1;1/2) 
  = 2^{(t+n)/2}J_{(t+n)/2}^{(0,-n-1)}(0).
\end{multline}
The results for the first sum are now proved.

The second sum, say $T$, can be treated in much the same way since
\begin{multline}
 T=(l-t) \;_{2}F_{1}(1-l,l-t+1;2;-1) =2^{l-1}(l-t) 
  \;_{2}F_{1}(1-l,2-l+t-1;2;1/2)\\
  =2^{(t-n)/2-1}(l-t) \;_{2}F_{1}(-(l-1),2+l-1+n;2;1/2) \\
  =(l-t)2^{(t-n)/2-1}J_{(t-n)/2-1}^{(1,n)}(0)/l.
\end{multline}
The result for nonnegative $n$ now follows.  If $n$ is negative
\begin{multline}
 T= \;_{2}F_{1}(-(n+t)/2+1,1-l;2;-1)\frac{t+n}{2} \\
  =\frac{t+n}{2}2^{(t+n)/2-1} \;_{2}F_{1}(-(t+n)/2-1,2+(n+t)/2-1-;2;1/2)\\
  =\frac{t+n}{2}2^{(t+n)/2-1} 
\;_{2}F_{1}(-((t+n)/2-1),2+((n+t)/2-1)-t;2;1/2)\\
=\frac{t+n}{t+n}2^{(t+n)/2-1}J_{(t+n)/2-1}^{(1,-n)}(0).
\end{multline}
This completes the proof of Lemma~\ref{second}.

We consider first of all $\psi_{L}(n,t),\; n \ge 0.$ (As was observed by
Ambainis {\emph{et al.,}} there is a symmetry between positive and 
negative $n$.)  We let 
\begin{equation}
  m=(1-\alpha)t/2-1
\end{equation}
so that
\begin{equation}
  1+n=1+\alpha t=\frac{1+\alpha}{1-\alpha}+\frac{2\alpha}{1-\alpha} m
\end{equation}
and
\begin{equation}
J_{(1-\alpha)t/2-1}^{(0,n+1)}(0) = 
J_{m}^{(0+0\cdot m, \frac{1+\alpha}{1-\alpha}+\frac{2\alpha m}{1-\alpha})}(0).
\end{equation}

\subsection{The oscillatory range:$|\alpha| < 1/\sqrt{2}-\varepsilon$}

For $0\le \alpha <2^{-1/2}-\varepsilon, \quad 0<\varepsilon \le 2^{-1/2},$ 
we may use either the Chen-Ismail \cite{ChenIsmail} results or the
Gawronkski-Shawyer \cite{Gawyer} results.  We use the results by 
Gawronkski and Shawyer since they have been proved to hold uniformly over 
this range of $\alpha.$  For the sake of consistency of notation and ease 
of reference we will state the Gawronkski and Shawyer result in a more 
restricted form than they obtained in their paper, but it is sufficient 
for our purposes.

Gawronkski and Shawyer write, using the integral representation in equation 
(4.46) of Szeg{\"{o}}'s book \cite{Szego},
\begin{equation}
 J_{m}^{(am+\gamma,bm+\beta)}(z) =\frac{1}{2\pi i} 
        \int_{\Gamma}e^{mh(\zeta)}g(\zeta) d\zeta
\end{equation}
where 
\begin{align}
 h(\zeta)&=\ln\left(\frac{\zeta^{2}-1}{2(\zeta-z)}\right)
  +a\ln\left(\frac{1-\zeta}{1-z}\right)
  +b\ln\left(\frac{1+\zeta}{1+z}\right) \\
  g(\zeta)&=\left(\frac{1-\zeta}{1-z}\right)^{\gamma}
  \left(\frac{1+\zeta}{1+z}\right)^{\beta}
  \frac{1}{\zeta-z} 
\end{align} 
and $\Gamma$ is a contour circling the origin.  These integrals are of the 
form
\begin{equation}\label{steepdef1}
 I(m)=\int_{\Gamma} g(\zeta)e^{-m h(\zeta)}\;d\zeta
\end{equation}
and as such can be approximated in the limit as $m \to \infty$ using the 
method of steepest descents \cite{Arfken,Olver,Wong}.  This relies the fact 
that in this limit (i.e., $t \to \infty$) the only parts of the integrand 
that contribute significantly to the integral are those regions where the 
function in the exponent, $h,$ has a maximum.  This is because in the long 
time limit the exponential term in the integrand behaves more and more like 
$\delta$-function(s) centred on point(s) where $h$ is maximal.  (Note that 
a stationary point of $h$ is only a maximum along a given contour of 
integration, $\Gamma.$  This is because the stationary points of an analytic 
function can only be saddle-points, so whether they appear to be maxima or 
minima depends on the path taken through them.)

In order to make use of this phenomenon, we need to be able to assume that the 
imaginary part of $h$ is approximately constant in the vicinity of these 
saddle-points, otherwise the integrand will oscillate unmanageably in the 
asymptotic limit.  (We don't care if it oscillates wildly elsewhere, as the 
contribution from those regions will be negligibly small.)  The way to 
achieve this is to choose the contour so that it passes through these 
saddle-points along the path of steepest descent for the real part of the 
exponent. 

\begin{figure}[floatfix]
 \begin{minipage}{\columnwidth}
  \begin{center}
   \resizebox{0.8\columnwidth}{!}{\includegraphics{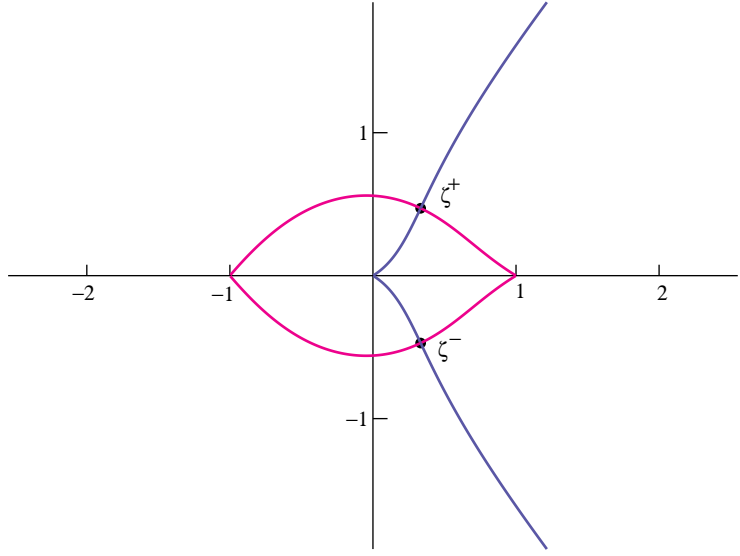}}
  \end{center}
 \end{minipage}
\caption{Steepest descent and ascent curves for the oscillatory range 
         of  $\alpha.$ The descent curve is the leaf-shape, and the two 
         lines running off to $\pm \infty$ are the steepest ascent 
         curves.}\label{leaf}
\end{figure}

We therefore choose the path $\Gamma$ so it goes through the two saddle-points 
$\zeta^{+},\zeta^{-}$ determined by
\begin{equation}
 h'(\zeta)=0(\text{ or }(1+\alpha)\zeta^{2}-2\alpha \zeta +1-\alpha=0).
\end{equation}
Thus the imaginary part of $h(\zeta)$ is fixed which implies that 
the real part of $h(\zeta)$ has a maximum at the saddle-point. 
(The numbers $e^{h(\zeta^{+})}, e^{h(\zeta^{-})}$ are the reciprocals of
the singularities found by Chen and Ismail using Darboux's method.)  
Gawronkski and Shawyer use a steepest descent contour which goes through
the saddle-points at $\zeta =\zeta^{+}$ or $\zeta^{-}$ and the
points $\zeta=\pm 1.$ The contour must be modified slightly near the 
singularities at $\zeta =\pm 1.$  The contours are leaf-shapes defined by
\begin{equation}
 \Im (h(\zeta))=\Im (h(\zeta^{+})).
\end{equation}
So (for example) when $\alpha =1/2$ this is \begin{equation}
\Im(\ln(\zeta^{2}-1)-\ln 2-\ln \zeta +2\ln(1+\zeta)) =\Im
(\ln((\zeta^{+})^{2}-1)-\ln 2 -\ln \zeta^{+}+2\ln(1+\zeta^{+}))
\end{equation}
where
\begin{equation}
 \zeta^{+}=\frac{\alpha+i \sqrt{1-2\alpha^{2}}}{1+\alpha}
 =\frac{1+i\sqrt{2}}{3}.
\end{equation}
Figure \ref{leaf} is an example of the steepest descent curve for the 
oscillatory range of $\alpha.$

The Gawronkski and Shawyer result is stated in terms of the parameters
$A,B,D,C_{1},C_{2},\theta,\rho(\theta)$ and $\xi(\theta)$ defined as 
follows:
\begin{align}
  &\cos \theta=-\frac{\alpha^{2}}{1-\alpha^{2}}\quad \quad
   \text{ so } \quad \frac{\pi}{2} \le \theta <\pi, \\
  &A=0, \quad \quad B=\alpha, \quad \quad D=1-\alpha^{2}, \\
  &C_{1}=1, \quad C_{2}=\frac{1+\alpha}{1-\alpha}.
\end{align}
They then show that $\rho(\theta)$ increases monotonically
from $0$ to $\pi$ as $\theta$ increases from $0$ to $\pi$ or equivalently
as $|\alpha|$ increases from $0$ to $1/\sqrt{2}$.  Furthermore
\begin{align}
  \rho(\theta)&=\theta +\frac{2\alpha}{1-\alpha}\left(\frac{\theta}{2}-
\arctan\left(\frac{1+B^{2}-D}{2B}\tan \theta /2\right)\right), \\
\xi(\theta)&=\frac{\theta}{2}+\frac{\pi}{4}+\frac{1+\alpha}{1-\alpha}
\left(\frac{\theta}{2}-\arctan\left(\frac{1+B^{2}-D}{2B}
\tan \theta/2\right)\right).
\end{align}

Here $-\pi/2 < \arctan u <\pi/2$ for real $u$.

\begin{theorem}[Gawronkski and Shawyer \cite{Gawyer}]\label{GSthm1} 
\begin{multline}\label{GS1}
  J_{m}^{(0+0\cdot m,\frac{1+\alpha}{1-\alpha}
      +\frac{2\alpha m}{1-\alpha})}(0)
    =\left(\pi m(1+B)\sin \theta/2 \cos \theta /2\right)^{-1/2} \times \\
\left(C_{2}^{-1}(1+B)\left((C_{2}-1)^{2}/4+
  C_{2}\cos^{2}\theta/2\right)^{1/2}\right)^{-\frac{2\alpha m}{1-\alpha}
   -\frac{1+\alpha}{1-\alpha}} \left(\sin
   (m\rho(\theta)+\xi(\theta))+O(1/m)\right)
\end{multline}
as $m$ (i.e., $t$) $\rightarrow \infty$.
\end{theorem}

Now, by direct substitution,
\begin{align}
   \left(\pi m(1+B)\sin(\theta /2)\cos(\theta /2\right)^{-1/2}
 &=2\left(\frac{1}{\pi t \sqrt{1-2\alpha^{2}}}\right)^{1/2}\label{GS+2} \\
   C_{2}^{-1}(1+B)&=1-\alpha\label{GS+3} \\
   \left((C_{2}-1)^{2}/4+C_{2}\cos^{2}(\theta /2)\right)^{1/2}
   &=\frac{1}{\sqrt{2}(1-\alpha)}\label{GS+4}
\end{align}
Thus from equations \eqref{GS1},\eqref{GS+2},\eqref{GS+3} and \eqref{GS+4},
\begin{multline}
	J_{m}^{(0,2\alpha/(1-\alpha)+(1+\alpha)/(1-\alpha))}(0)=
  2\left(\frac{1}{\pi t \sqrt{1-2\alpha{^2}}}\right)^{1/2} \times \\
  2^{(2\alpha((1-\alpha)t-1))/(1-\alpha)+(1+\alpha)/(1-\alpha))/2}
  \sin\left((1-\alpha)t/2-1)\rho(\theta)+\xi(\theta)\right)(1+O(1/t)).
\end{multline}
We have that
\begin{equation}
  \alpha t-\frac{\alpha}{1-\alpha}+\frac{1+\alpha}{2(1-\alpha)}
   =\frac{\alpha t}{2}+\frac{1}{2}
\end{equation}
so
\begin{equation}
 J_{m}^{(0,\frac{2\alpha m}{1-\alpha}+\frac{1+\alpha}{1-\alpha})}(0)
=\frac{2\sqrt{2}\;2^{n/2}}{(\pi t \sqrt{1-2\alpha^{2}})^{1/2}} \cdot
\sin((1-\alpha)t/2-1)\rho(\theta)+\xi(\theta))(1+O(1/t)).
\end{equation}
Now it follows that for $-t \le n \le t,$
\begin{equation}
\psi_{L}(n,t)=\frac{\sqrt{2}}{(\pi t
  \sqrt{1-2\alpha^{2}})^{1/2}}
  \sin((1-\alpha)t/2-1)\rho(\theta)+\xi_{L}(\theta))(1+O(1/t))
\end{equation}
where
\begin{align}
 \rho(\theta)&=\frac{\theta}{1-\alpha}-\frac{2\alpha}{1-\alpha}
 \arctan\left(\frac{\alpha}{\sqrt{1-\alpha^{2}}}\right)\label{rhothe1} \\
\xi_{L}(\theta)&=\frac{\theta}{1-\alpha}+\pi/4 -\frac{1+\alpha}{1-\alpha}
\arctan\left(\frac{\alpha}{\sqrt{1-\alpha^{2}}}\right).\label{xi1}
\end{align}

When we consider $\psi_{\rm{R}}(n,t),\; 0 \le n \le t$ we find that the 
power of $2$ does not change.  (Gawronkski and Shawyer use the symbol 
$\alpha$ as a dummy variable in their Jacobi polynomials.  To avoid 
confusion, we will call this parameter $\kappa.$)  For  
$\psi_{\rm{L}}(n,t),\;\kappa$ was set to be zero.  To do the calculation 
for $\psi_{\rm{R}}(n,t)$ we need to set $\kappa=1.$  Note also that we 
must reset $\tau$ (which they call $\beta$) to be  
$\tau=b=2\alpha/(1-\alpha).$  Thus $\rho(\theta)$ does not change.  
However, $\xi(\theta)$ does change as we will specify below.  We (as 
Gawronkski and Shawyer do) use
\begin{equation}
 \frac{\pi}{2}=\lim_{A\rightarrow 0+}\left(\frac{1+A^{2}-B^{2}+D}{2A}
  \tan\frac{\theta}{2}\right)
\end{equation}
to obtain
\begin{equation}\label{psiR1}
 \zeta_{\rm{R}}(\theta)=\zeta_{\rm{L}}(\theta)-\frac{\pi}{2}+
 \arctan\left(\frac{\alpha}{\sqrt{1-\alpha^{2}}}\right).
\end{equation}

Thus we have:
\begin{theorem}\label{thmone}
Let $\varepsilon > 0$ be any constant and $\alpha$ be in the
interval $(-1/\sqrt{2}+\varepsilon,1/\sqrt{2}-\varepsilon).$ Then as $t
\rightarrow \infty$ we have uniformly for $0\le  
|\alpha|<1/\sqrt{2}-\varepsilon,$
\begin{align}  
  \psi_{\rm{R}}(n,t)&=\left(\frac{1+\alpha}{1-\alpha}\right)^{1/2}
    \frac{\sqrt{2}}{(\pi t \sqrt{1-2\alpha^{2}})^{1/2}}
    \sin\left((\frac{1-\alpha}{2}t-1)\rho(\theta)+
    \xi_{\rm{R}}(\theta)\right)(1+O(1/t)) \\
  \psi_{\rm{L}}(n,t)&=\frac{\sqrt{2}}{(\pi t \sqrt{1-2\alpha^{2}})^{1/2}}
    \sin\left((\frac{1-\alpha}{2}t-1)\rho(\theta)+
    \xi_{\rm{L}}(\theta)\right)(1+O(1/t))
\end{align}
with $\rho ,\xi_{\rm{L}}$ and $\xi_{\rm{R}}$ defined by equations 
\eqref{rhothe1}, \eqref{xi1} and \eqref{psiR1}.  Here
\begin{equation}
\theta =\arccos(-\alpha^{2}/(1-\alpha^{2})), \quad
 \frac{\pi}{2} \le \theta < \pi.
\end{equation}
\end{theorem}

{\emph{Remark:}} The term of the form $1/\sqrt{1-2\alpha^2}$ gives us 
forewarning that this term is going to become very large when 
$|\alpha|=1/\sqrt{2}.$ This is consistent with the graph in figure 
\ref{walk}.  In fact this term actually diverges at this value of 
$\alpha,$ but this is a symptom of the breakdown of this approximation, 
which is why we included the $\varepsilon$ in the statement of the 
theorem.  In this transitional range, Theorem \ref{SVthm1} below is the 
appropriate form to use.

\bigskip

Our Theorem \ref{thmone} agrees with Theorem 2 of Ambainis {\emph{et 
al.,}} as expected, and it also gives an estimate for the error term.  We 
skip the proof that the answers are identical for the $\psi$-functions as 
the probability function $p(n,t)$ defined by
\begin{equation}
 p(n,t)=\psi_{\rm{L}}^{2}(n,t)+\psi_{\rm{R}}^{2}(n,t)
\end{equation}
is more interesting and we will show that for this function and its 
moments, our answers are identical.  We will use the identity 
$\sin^{2}A=(1-\cos 2A)/2.$ Recall that $\alpha =n/t$, where we temporarily 
think of $t$ as fixed and let $n$ vary.  We may then use $B(n,t)$ to 
denote a bounded function with bounded derivatives and $A(n,t)$ for a 
function such that it and its derivatives are bounded away from $0,$ thus
for $\psi_{\rm{R}}(n,t)$ we have
\begin{align}
 A(n,t) &= \frac{1-\alpha}{2}\rho(\theta)+
           \frac{\xi_{\rm{R}}(\theta)-\rho(\theta)}{t} \\
 B(n,t) &= \left(\frac{1+\alpha}{1-\alpha}\right)^{1/2}
           \frac{\sqrt{2}}{(\pi t \sqrt{1-2\alpha^{2}})^{1/2}}
\end{align}
and for $\psi_{\rm{L}}(n,t)$ we have instead 
\begin{align}
 A(n,t) &= \frac{1-\alpha}{2}\rho(\theta)+ 
           \frac{\xi_{\rm{L}}(\theta)-\rho(\theta)}{t} \\
 B(n,t) &= \frac{\sqrt{2}}{(\pi t \sqrt{1-2\alpha^{2}})^{1/2}}.
\end{align}
This enables us to write:
\begin{multline}
 \int_{\alpha_{1}}^{\alpha_{2}} B(n,t)\cos(A(n,t)t) \; d\alpha = \\ 
 \int_{\alpha_1}^{\alpha_2} \left(B(n,t)
     \frac{d}{d\alpha}\left(\frac{\sin(A(n,t)t)}{tA'(n,t)}\right)
     + B(n,t)\frac{\sin(A(n,t)t)A''(n,t)}{t (A'(n,t))^2} \right)\;d\alpha = \\
 \int_{\alpha_1}^{\alpha_2} B(n,t)
      \frac{d}{d\alpha}\left(\frac{sin(A(n,t)t)}{tA'(n,t)}\right) 
     \;d \alpha + O(1/t) = \\
 \left[\frac{B(n,t)\sin(A(n,t)t)}{A'(n,t)t}\right]_{n_{1},t_{1}}^{n_{2},t_{2}} 
  -\int_{n_{1},t_{1}}^{n_{2},t_{2}} 
   \frac{B'(n,t)\sin(A(n,t)t)}{A'(n,t)t}\; d\alpha +O(1/t) = O(1/t)
\end{multline}
by a simple integration by parts, where 
$\alpha_1=n_1/t_1,\; \alpha_2=n_2/t_2.$ If we write $p(\alpha)=tp(n,t)$ 
(following Ambainis {\emph{et al.}}) and note that the $O$-term is uniform
for $-1/\sqrt{2}+\varepsilon <\alpha < 1/\sqrt{2}-\varepsilon,$ then that
observation and our Theorem \ref{thmone} give us that 
\begin{equation}
 \int_{\alpha_{1}}^{\alpha_{2}}p(\alpha) \; d\alpha
  =\int_{\alpha_{1}}^{\alpha_{2}}\left(\frac{1}{\pi\sqrt{1-2\alpha^{2}}}
  +\left(\frac{1+\alpha}{1-\alpha}\right)\frac{1}{\pi\sqrt{1-2\alpha^{2}}}
  \right)\;d\alpha+O(1/t)
\end{equation}
provided $|\alpha|<1/\sqrt{2}-\varepsilon$. Note for the quantum walk
$p(\alpha)$ is 0 when $n$ and $t$ have unequal parity so for the quantum
walk we have
\begin{equation}
\int_{\alpha_{1}}^{\alpha_{2}}p(\alpha)\;d \alpha
 =\frac{1}{\pi}\int_{\alpha_{1}}^{\alpha_{2}}
  \frac{1}{(1-\alpha)\sqrt{1-2\alpha^{2}}} \;d\alpha +O(1/t)
\end{equation}
To confirm that we have a probability distribution, we must verify that 
the function integrates to $1:$
\begin{align}
 \int_{\varepsilon-1/\sqrt{2}}^{-\varepsilon+1/\sqrt{2}}p(\alpha)\;d\alpha
   &\sim 
 \frac{1}{\pi}\int_{\varepsilon-1/\sqrt{2}}^{-\varepsilon+1/\sqrt{2}}
 \frac{1}{(1-\alpha)\sqrt{1-2\alpha^{2}}}\;d\alpha \\
   &\sim
\frac{1}{\pi}\int_{\varepsilon-1/\sqrt{2}}^{-\varepsilon+1/\sqrt{2}}
\frac{1}{(1-\alpha^{2})\sqrt{1-2\alpha^{2}}}\;d\alpha.
\end{align}
If we let $\cos k=-\alpha/\sqrt{1-\alpha^{2}},$ and 
$\sin k = \sqrt{1-2\alpha^{2}}/\sqrt{1-\alpha^{2}},$ then we can write
\begin{equation}
 \frac{dk}{d\alpha}=\frac{1}{(1-\alpha^{2})\sqrt{1-2\alpha^{2}}}
\end{equation}
so
\begin{equation}
  \frac{1}{\pi}\int_{\varepsilon-1/\sqrt{2}}^{-\varepsilon+1/\sqrt{2}}
  p(\alpha)\;d\alpha 
=\frac{1}{\pi} \int_{-\pi+\delta(\varepsilon)}^{\delta(-\varepsilon)}\;dk
=1-2\pi^{-1}\delta(\varepsilon)
\end{equation}
where ${\lim}_{\varepsilon \rightarrow 0} \;\delta(\varepsilon)=0,$ as 
required.  The correction term appears because we have only performed the 
integration over the oscillatory range of the probability function, as 
this is where it has almost all of its support. 

\bigskip

We can now write down the $m$-th moment of the distribution:
\begin{equation}
 \sim\frac{1}{\pi}\int_{-1/\sqrt{2}}^{1/\sqrt{2}}
 \frac{\alpha^{m}}{(1-\alpha)\sqrt{1-2\alpha^{2}}}\;d\alpha.
\end{equation}
Thus the first moment is
\begin{multline}
\frac{1}{\pi} \int_{-1/\sqrt{2}}^{1/\sqrt{2}}
\frac{\alpha^{2}-1}{(1-\alpha^{2})\sqrt{1-2\alpha^{2}}}\;d\alpha
+\frac{1}{\pi}\int_{-1/\sqrt{2}}^{1/\sqrt{2}}\frac{1}{(1-\alpha)
\sqrt{1-2\alpha^{2}}}\;d\alpha \\
=-\frac{1}{\pi}\int_{-1/\sqrt{2}}^{1/\sqrt{2}}
\frac{1}{\sqrt{1-2\alpha^{2}}}\;d\alpha + 1=1-1/\sqrt{2}.
\end{multline}
The second moment can be seen to be also equal to 1-$1/\sqrt{2}$.

\subsection{The exponential range: $|\alpha| > 2/\sqrt{2}+\varepsilon$}

We now consider the range $\alpha \ge 1/\sqrt{2}+\varepsilon$ where
$0<\epsilon < 1-1/\sqrt{2}$. The Gawronkski and Shawyer results are an
extension and refinement of the results of Saff and Varga \cite{Saff}. We
state the results as Saff and Varga do.  They write
\begin{equation}
  J_{m}^{(\gamma,2\alpha m/(1-\alpha)+\beta)}(0)=
   \int_{\Gamma}e^{mh(\zeta)}g(\zeta)d\zeta
\end{equation}
where
\begin{align}
  h(\zeta)&=\ln(\zeta^{2}-1)-\ln 2-\ln\zeta+ 
    \frac{2\alpha}{1-\alpha}\ln(1+\zeta), \\
g(\zeta)&=\frac{1}{2\pi i}(1+\zeta)^{\beta}(1-\zeta)^{\gamma} 
\frac{1}{\zeta}.
\end{align}
They choose $\zeta$ to be the saddle-point $\zeta^{-},$ where
\begin{equation}
\zeta^{-}=\frac{\alpha-\sqrt{2\alpha^{2}-1}}{1+\alpha}.
\end{equation}
Using the saddlepoint method they derive
\begin{theorem}[Saff and Varga \cite{Saff}]\label{SVthm1}
\begin{equation}
  J_{m}^{(\gamma,2\alpha m/(1+\alpha)+\beta}(0)\sim i\exp[mh(\zeta^{-})]
      g(\zeta^{-})\left(\frac{2\pi}{mh''(\zeta^{-})}\right)^{1/2}
\end{equation}
where
\begin{equation}\label{hdprime1}
  h^{''}(\zeta)=-2\frac{\zeta^{2}+1}{(\zeta^{2}-1)^{2}}
   +\frac{1}{\zeta^{2}}-\frac{2\alpha}{(1-\alpha)(1+\zeta^{2})}.
\end{equation}
\end{theorem}
Now
\begin{align}
 \zeta-\zeta^{-1}&=\frac{\alpha-\sqrt{2\alpha^{2}-1}}{1+\alpha}
  -\frac{\alpha+\sqrt{2\alpha^{2}-1}}{1-\alpha}
  =-2\frac{\alpha^{2}+2\sqrt{2\alpha^{2}-1}}{1-\alpha^{2}},\\
 1+\zeta &=\frac{1+2\alpha-\sqrt{2\alpha^{2}-1}}{1+\alpha}.
\end{align}
Then according to Saff and Varga,
\begin{multline}
  J_{m}^{(\gamma,2\alpha m/(1-\alpha)+\beta}(0)
    \sim \left(\frac{\alpha^{2}+\sqrt{2\alpha^{2}-1}}
    {1-\alpha^{2}}\right)^{m}\left(\frac{1+2\alpha
    -\sqrt{2\alpha^{2}-1}}{1+\alpha}\right)^{2\alpha m/(1-\alpha)} 
\times\\ 
   \frac{1}{2\pi} 
\left(\frac{1+2\alpha-\sqrt{2\alpha^{2}-1}}{1+\alpha}\right)^{\beta}
\left(\frac{1+\sqrt{2\alpha^2 -1}}{1+\alpha} \right)^{\gamma}
\frac{1+\alpha}{\alpha-\sqrt{2\alpha^{2}-1}}
\left(\frac{2 \pi}{mh^{''}(\zeta)}\right)^{1/2}
\end{multline}
where $h^{''}(\zeta)$ is defined in \eqref{hdprime1}.

Gawronkski and Shawyer show that the $\sim$ symbol can be replaced by
$1+O(1/t)$ and that this expansion holds uniformly for $\alpha \in 
[1/\sqrt{2}+\epsilon,1-\epsilon].$

Note that when $\alpha \rightarrow 1/\sqrt{2},$
\begin{equation}
  \left|J_{m}^{(\gamma,2\alpha m/(1-\alpha)+\beta)}(0)\right|^{1/m}
   \rightarrow \left(\frac{1+\sqrt{2}}{1+1/\sqrt{2}}\right)^
   {(\frac{2\alpha}{1-\alpha})(\frac{1-\alpha}{2}t)}
  =2^{\alpha t/2}=2^{n/2}
\end{equation}
so the asymptotic estimate above metamorphoses into the asymptotic 
estimate in Theorem \ref{thmone}.  We shall state our results for positive 
$\alpha$ as those for negative $\alpha$ follow immediately.  These results 
refine the estimates of Ambainis {\emph{et al.}} \cite{Ambainis01}.
\begin{theorem}\label{thmtwo}
  If $\alpha \ge 1/\sqrt{2}+\varepsilon$, then uniformly for
$\alpha \in [1/\sqrt{2} + \varepsilon,1-\varepsilon],$
\begin{align}
 \psi_{\rm{R}}(n,t)&=C_{\rm{R}}\frac{1}{t^{1/2}}B^{t}(\alpha)(1+O(1/t))\\
 \psi_{\rm{L}}(n,t)&=C_{\rm{L}}\frac{1}{t^{1/2}}B^{t}(\alpha)(1+O(1/t))
\end{align}
where $B(\alpha), \; C_{\rm{L}}, \; C_{\rm{R}}$ are defined in the  
asymptotic expansion of 
\begin{equation}
  2^{-n/2-1}J_{m}^{(\gamma,2\alpha m/(1-\alpha)+\beta)}(0)
\end{equation}
following from the above. (Note that for $C_{\rm{L}}, \; \gamma=0$ and
$\beta=(1+\alpha)/(1-\alpha)$ while for $C_{\rm{R}}, \; \gamma=1, \;
\beta=2\alpha/(1-\alpha)$ in the above.)  Thus 
\begin{equation}
 B(\alpha) = \left(\frac{1+2\alpha - \sqrt{2\alpha^2 -1}}{1+\alpha}
                  \right)^{\alpha} 
             \left(\frac{\alpha^2+\sqrt{2\alpha^2-1}}{1-\alpha^2}
                  \right)^{\frac{1-\alpha}{2}}
\end{equation}
and
\begin{equation}
 C_{\rm{R}}=\left(\frac{1+\sqrt{2\alpha^2-1}}{1+2\alpha-\sqrt{2\alpha^2-1}}
            \right) \times C_{\rm{L}}(\alpha)
\end{equation}
where
\begin{multline}
 C_{\rm{L}}(\alpha) = \frac{1}{\sqrt{2\pi}} 
                      \left(\frac{1+2\alpha-\sqrt{2\alpha^2-1}}{1+\alpha})
                           \right)^{\frac{1+\alpha}{1-\alpha}}
                      \left(\frac{1+\alpha}{\alpha-\sqrt{2\alpha^2-1}}\right)
                      \times \frac{2}{1-\alpha} \times \\
   \left(\left(\frac{\alpha+\sqrt{2\alpha^2-1}}{1-\alpha}\right)^2
      -\frac{(1+\alpha)^2}{(1-\alpha)(2\alpha+1-\sqrt{2\alpha^2-1})}\right. \\
      \left.- \frac{\alpha(\sqrt{2\alpha^2-1} - (1+2\alpha))(1+\alpha)^2}
        {(1+\alpha(1-\alpha+\sqrt{2\alpha^2-1}))^2}
        \right)^{-1/2}
\end{multline}
to reconstruct the complete wavefunction.
The asymptotics for $\alpha$ in the range 
$-1+\epsilon \le \alpha \le -1/\sqrt{2}-\varepsilon$ follow from the spatial 
symmetry between $-n$ and $n$.
\end{theorem}

\begin{figure}[floatfix]
 \begin{minipage}{\columnwidth}
  \begin{center}
   \resizebox{0.8\columnwidth}{!}{\includegraphics{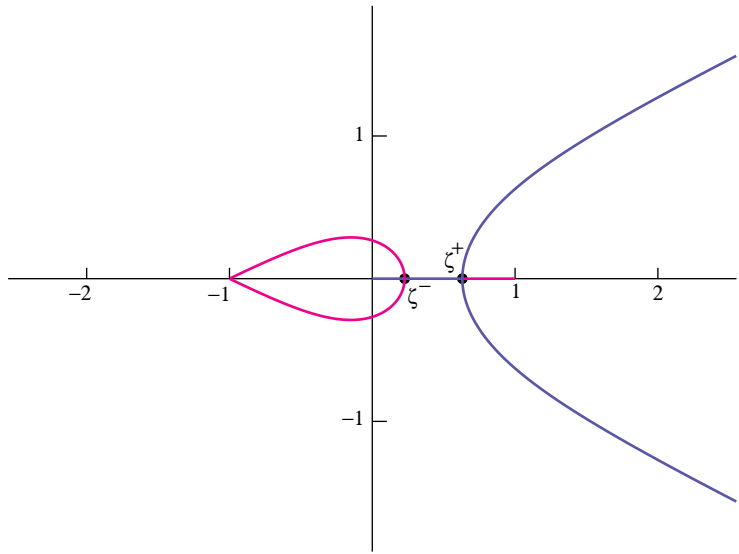}}
  \end{center}
 \end{minipage}
\caption{Steepest descent and ascent curves for $|\alpha|>1/\sqrt{2}.$ The 
teardrop shape and the line from $\zeta^+$ are both steepest  descent 
curves, but only the contour that surrounds the origin  gives a valid 
integral.  The remainder of the lines are steepest ascent curves.}
\label{evanesce}
\end{figure}

Figure \ref{evanesce} shows the steepest descent curves for this
range.

\bigskip

{\emph{Remark:}} 
Saff and Varga show that $B(\alpha)$ is a decreasing function of 
$\alpha$ for $\alpha \in [1/\sqrt{2}+\varepsilon,1-\varepsilon].$ Thus the 
$\psi-$ functions decrease exponentially in $t$ in this range.

\subsection{The transitional range: $2^{-1/2}-\varepsilon < \alpha < 
2^{-1/2} + \varepsilon$}

We now consider the range $1/\sqrt{2}-\varepsilon<\alpha <1/\sqrt{2}+
\varepsilon$.  Theorems \ref{thmone} and \ref{thmtwo} exhibit an 
oscillating sine term times $t^{-1/2}$ for 
$|\alpha|<1/\sqrt{2}-\varepsilon$ and an exponentially small estimate for
$|\alpha|>1/\sqrt{2}+\varepsilon.$ Saff and Varga \cite{Saff} show that if
$\alpha =1/\sqrt{2}$ then
\begin{equation}
 J_{m}^{(\gamma,2\alpha/(1-\alpha)+\beta)}(0)
 \sim\frac{g(2\alpha-1)}{3}\left(\frac{6}{mh^{(3)}(2\alpha-1)}\right)^{1/3}
  e^{i\pi /3}\Gamma(1/3)
\end{equation}
(See also Ambainis-et al, \cite{Ambainis01}.) The asymptotic behaviour is  
therefore qualitatively different in the three cases 
$|\alpha|<1/\sqrt{2}-\varepsilon, \quad \alpha=1/\sqrt{2},$ and $\quad 
|\alpha|>1/\sqrt{2}+\varepsilon.$

We now apply the analysis for coalescing saddle-points as described in R. 
Wong's book \cite{Wong}, to get a uniform asymptotic expansion for a range 
of $\alpha$ extending a positive distance on each side of 
$\alpha=1/\sqrt{2}$.  The asymptotic behaviour is described in terms
of the Airy function, $Ai(x).$ These functions cannot be written out 
explicitly, but (like Bessel functions) they can be written in terms
of integral representations that are well understood.  It suffices for
us to know that as $x \rightarrow -\infty$ the behaviour of these
functions is oscillatory,
\begin{align}
  Ai(-x) &\sim \frac{1}{\sqrt{\pi}}\cos(\frac{2}{3}x^{3/2}-\pi/4)/x^{1/4}, \\
  Ai'(-x)&\sim \frac{x^{1/4}}{\sqrt{\pi}}\sin(\frac{2}{3}x^{3/2}-\pi/4)
\end{align}
but that when $x \rightarrow \infty$ they behave like
\begin{align}
  Ai(x) &\sim \frac{1}{2\sqrt{\pi}}\pi x^{1/4}  
           \exp\left(-\frac{2}{3}x^{3/2}\right), \\
  Ai'(x) &\sim -\frac{1}{2\sqrt{\pi}}x^{1/4}
           \exp\left(-\frac{2}{3}x^{3/2}\right)
\end{align}
which decreases faster than any power of $x$.

Recall that we have the two saddle-points
$\zeta^{-},\zeta^{+}=(\alpha \mp \sqrt{2\alpha^{2}-1})/(1+\alpha)$.
Following Wong we define the variables $\zeta$ and $\eta$ by
\begin{align}
  \zeta^{3/2}&=\frac{3}{4}(h(\zeta^{+})-h(\zeta^{-})) \\
   \text{ or }\quad  
2\zeta^{1/2}=\zeta^{+}-\zeta^{-}&=2i\sqrt{1-2\alpha^{2}}/(1+\alpha),\\
\eta&=\frac{1}{2}(h(\zeta^{-})+h(\zeta^{+}))
\end{align}

We must choose $\zeta^{-},\zeta^{+}$ as we have because we want the range 
of $\alpha > 1/\sqrt{2}$ to give positive $\zeta$ so that there is 
exponential decay as $\zeta \rightarrow \infty$.  Note also (as Saff and 
Varga point out) that if $\alpha =1/\sqrt{2}$ then
\begin{align}
 h^{'}(\alpha)&=h^{''}(\alpha)=0 \\
 h^{'''}(\alpha)&=-\frac{1+\alpha}{24\alpha^{2}(1-\alpha)^{3}}.
\end{align}
Thus
\begin{equation}
 h(1/\sqrt{2}+\delta)
  =-\frac{1+\alpha}{24\alpha^{2}(1-\alpha)^{3}}\delta^{3}+O(\delta^{4})
\end{equation}
so
\begin{equation}
  mh(\alpha)=o(1) \quad \text{ if } \quad \delta =o(m^{-1/3}).
\end{equation}
Also $h(\zeta^{+})\sim h(\zeta^{-})$ so $\eta \sim h(\zeta^{-})$
for this range of $\alpha$.

Suppose for now that equation 4.31 in chapter VII of Wong \cite{Wong} 
holds, which in our notation becomes
\begin{equation}
  J_{m}^{(\gamma,2\alpha m/(1-\alpha)+\beta)}(0)
  =2\pi i e^{-m \eta}\left(\frac{Ai(m^{2/3}\zeta)}{m^{1/3}} 
  a_{0}+\frac{Ai^{'}(m^{2/3})}{m^{2/3}}
  b_{0}\right)(1+O(m^{-1/3}))
\end{equation}
where $a_{0}$ and $b_{0}$ are independent of $m.$ (In Wong's book, our $m$ 
is his $\lambda$ and our $\zeta$ is his $t.$ Our Jacobi polynomial is his 
$I(\lambda;\alpha),$ but his $\alpha$ is something else.)  To obtain an 
expression for $\psi_{\rm{L}}$ we need to set $\gamma=0$ and 
\begin{equation}
 \beta= \frac{1+\alpha}{1-\alpha}
\end{equation}
as in Theorem 
\ref{thmone}.  
Likewise, for $\psi_{\rm{R}},$ we will need to set $\gamma=1$ and
\begin{equation}
  \beta=\frac{2\alpha}{1-\alpha}.
\end{equation}
The previous argument shows that the term $m\eta =o(1)$ if $\delta
=o(m^{-1/3})$ and so the $e^{m\eta}$ term is asymptotic to $1.$ Note
furthermore that if $\alpha = 1/\sqrt{2}+\delta$ then the definition of
$\zeta$ implies that
\begin{equation}
  \zeta=-\frac{1-2\alpha^{2}}{1+\alpha^{2}} \quad \quad
  =\frac{-4\sqrt{2}}{(1+\sqrt{2})^{2}}\delta+O(\delta^{2}).
\end{equation}
Thus
\begin{equation}
 m^{2/3}\zeta
 =-\frac{4\sqrt{2}}{(1+\sqrt{2})^{2}}m^{2/3}\delta+O(\delta^{2}).
\end{equation}
When we use this estimate in the asymptotic behaviour of $Ai(m^{2/3}\zeta)$
and $Ai^{'}(m^{2/3}\zeta)$ we see that the Airy functions give terms
superpolynomially small if $\delta > m^{-2/3+\epsilon}.$ If $\delta
=m^{-\eta},\; \eta < 2/3$ however we see that the $\psi$-functions are
only polynomially small.  A similar argument shows that if $|\alpha| <
2^{-1/2}-m^{-2/3+\epsilon}$ then the behaviour of the $\psi$
is oscillatory.

We can therefore write that the transition from oscillatory behaviour 
bounded below by a power of $m$ or $t$ to bounded above by a 
superpolynomially small function occurs when $\alpha$ varies
by $O(t^{-2/3})$ from $1/\sqrt{2}$.

We may use equation 4.31 in chapter VII of Wong \cite{Wong} if the 
transformation
\begin{equation}
 h(\zeta)=u^{3}/3-\zeta u+\eta
\end{equation}
is single-valued on the contour of integration.  We may use the
Gawronkski-Shawyer contour of integration (which Saff and Varga also use).
The transformation will be one-to-one if on the path of integration,
\begin{equation}
  \frac{d\zeta}{du}\ne 0 \text{ and } \frac{du}{d\zeta}\ne 0.
\end{equation}
Now
\begin{equation}
 \frac{d\zeta}{du}=\frac{u^{2}-\zeta}{h^{'}(\zeta)}
\end{equation}
and the bottom derivative is only zero at the saddle-points.  The only 
saddle-points are at $\zeta^{-}$ and $\zeta^{+}$.  Wong shows that the
choice of $\zeta^{-},\zeta^{+}$ we used implies that $\frac{d\zeta}{du}\ne 
0.$ The only place the numerator is zero is at $\zeta^{-},\zeta^{+}$ and 
the denominator is analytic and therefore bounded on the contour of 
integration.  We find from Wong's book that

\begin{theorem}\label{thmthree}
There is a positive $\varepsilon$ so that uniformly for some $a_{0}$ and 
 $b_{0}$ (defined below)
\begin{equation}
 J_{m}^{(\gamma,2\alpha m/(1-\alpha)+\beta)}(0) =2\pi i e^{-m\eta}\left(
\frac{Ai(m^{2/3}\zeta)}{m^{1/3}}a_{0}+\frac{Ai'(m^{2/3}\zeta)}{m^{2/3}}
b_{0}\right)(1+O(m^{-1/3})).
\end{equation}
When $ n \equiv t$ mod 2 and $J_{r}^{(u,v)}(z)$ denotes a Jacobi 
polynomial,
\begin{equation}
 \psi_{L}(n,t) = 2^{-n/2-1}
  J_{m}^{(0,\frac{2\alpha m}{1-\alpha}+\frac{1+\alpha}{1-\alpha}}(0), 
  \quad \text{ when }0 \le n < t, 
\end{equation}
where we can use the Remark following lemma \ref{second} to 
obtain the wavefunction for negative $n.$ 
Also,
\begin{equation}
\psi_{R}(n,t) = (\frac{1+\alpha}{1-\alpha})2^{-n/2-1}
  J_{m}^{(1,\frac{2\alpha m}{1-\alpha}+\frac{2\alpha}{1-\alpha})}(0),
 \quad \text{ when }0\le n <t,
\end{equation}
and use the symmetry properties to obtain the other half as before. 
\end{theorem}
So this integral representation is valid for all values of $-1+\varepsilon 
\leq \alpha \leq 1-\varepsilon.$ In this theorem, 
\begin{equation}
  \phi_{0}(u)=g(\zeta)\frac{d\zeta}{du}
\end{equation}
and
\begin{align}
 a_{0}&=\frac{1}{2}[\phi_{0}(\zeta^{1/2})+\phi_{0}(-\zeta^{1/2})], \\
 b_{0}&=\frac{1}{2}[\phi_{0}(\zeta^{1/2})-\phi_{0}(-\zeta^{1/2})].
\end{align}

\bigskip

{\emph{Remark:}}
Theorem \ref{thmthree} gives the asymptotic behaviour in a rather
convoluted way.  It is really only useful for $|\alpha|$ very near to
$1/\sqrt{2},$ where the behaviour undergoes a qualitative change from
oscillatory to exponential decay.  For other values of $\alpha,$ Theorems
\ref{thmone} and \ref{thmtwo} give much simpler expressions for the
wavefunction. 

\bigskip

\begin{figure}[floatfix]
 \begin{minipage}{\columnwidth}
  \begin{center}
   \resizebox{0.8\columnwidth}{!}{\includegraphics{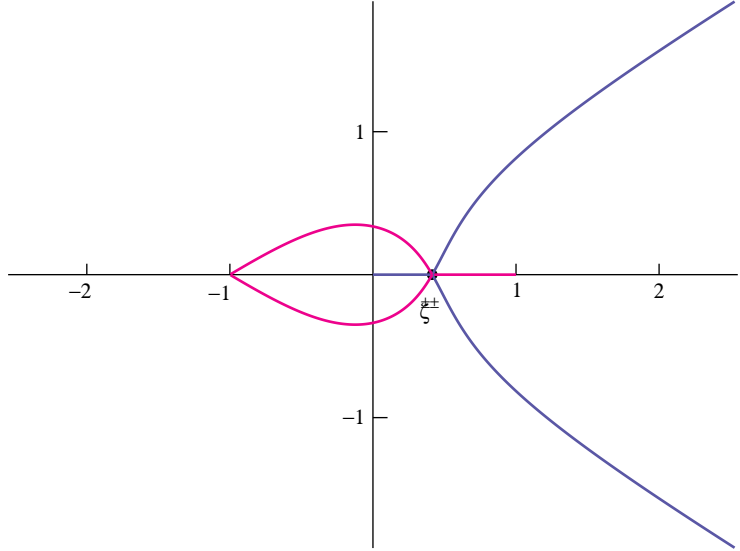}}
  \end{center}
 \end{minipage}
\caption{Steepest descent and ascent curves when the saddle-points 
coalesce.  The tear-drop shape and the line from $\zeta^{\pm}$ comprise 
the steepest descent curve, and the two lines out to $\pm\infty,$ 
together with the line to the origin are the steepest ascent curve.}
\label{drop}
\end{figure}

Figure \ref{drop} shows the steepest descent curve when the two 
saddle-points coalesce. 

\subsection{Error bounds for the method of steepest descents.}

We now discuss very briefly the error-terms in our results so far.  We
restrict our attention to $\psi_{L}(n,t)$ and Theorem \ref{thmone} above,
but similar comments apply to the other results in this paper and
$\psi_{R}(n,t)$.  In order to apply Theorem 7.1 of Olver we expand
$h(\zeta)$ and $g(\zeta)$ in powers of $\zeta-\zeta^{+}$:
\begin{align}
  e^{mh(\zeta)}&= e^{mh(\zeta^{+}) 
  +\sum_{r=3}^{\infty}b_{r}(\zeta-\zeta^{+})^{r}} \\
  g(\zeta)&=g(\zeta^{+})+\sum_{r=1}c_{r}(\zeta-\zeta^{+})^{r}
\end{align}
and write
\begin{equation}
 e^{\sum_{r=3}^{\infty}b_{r}(\zeta-\zeta^{+})^{r}}
\left(g(\zeta^{+})+\sum_{r=1}^{\infty}c_{r}(\zeta-\zeta^{+})^{r}\right)
=\sum_{s=0}^{\infty}a_{s}(\zeta-\zeta^{+})^{s}.
\end{equation}

The steepest descent contour that Gawronkski and Shawyer use naturally
separates into two pieces, a piece $\Gamma_{1}$ above the real axis and a
piece $\Gamma_{2}$ below the real axis.  Notice that $\Gamma_{1}$ is the
mirror image of $\Gamma_{2}$ in the real axis. With this notation, Theorem 
7.1 in Olver's book \cite{Olver} says
\begin{equation}
  \int_{\Gamma_{1}}e^{-mh(\zeta)}g(\zeta)
   \sim 2e^{-mh(\zeta^{+})}\sum_{s=0}^{\infty}
   \Gamma(s+1/2)\frac{a_{2s}}{m^{s+1/2}}.
\end{equation}
Here
\begin{align}
  a_{0}&=\frac{g}{(2h'')^{1/2}} \\
  a_{2}&=\left\{2g''-\frac{2h'''}{h''}
  +\left(\frac{5(h''')^{2}}{6(h'')^{2}}-\frac{h''''}{2h''}\right)g
  \right\}\frac{1}{(2h'')^{1/2}}
\end{align}
when $g,h$ and their derivatives are evaluated at $\zeta=\zeta^{+}.$ The
asymptotic expansion of the integral over $\Gamma_{2}$ is the complex
conjugate of the integral over $\Gamma_{1}$.  It is therefore possible to
derive complete asymptotic expansions for $\psi_{\rm{L}}(n,t)$ and 
$\psi_{\rm{R}}(n,t)$ and certainly Gawronkski and Shawyer were aware of 
this. The $s=0$ term is the asymptotic formula of Theorem \ref{thmone} 
above. Since we do not need more precision than in Theorem \ref{thmone} 
for this application, we do not pursue this further.

Olver then explains how to derive numerical estimates for the error term.
This is somewhat complicated so we refer the reader to section 10 of 
chapter 4 of Olver's book.  One needs to compute the maximum of certain 
quantities on the contour of integration.  Since we do not have 
applications of such bounds we will not pursue that here.  Olver also 
shows how to derive explicit numerical error bounds when the asymptotic 
expansion immediately above is truncated at any value of $s$.

\section{The Schr\"{o}dinger Approach}

Nayak and Vishwanath \cite{NayaknV} start from the recursion relations in 
\eqref{recursion} and use the Fourier transform
\begin{equation}
\tilde\Psi(k,t)=\sum_{n}\Psi(n,t)e^{ikn}
\end{equation}
where $\psi(n,t)$ is defined by \eqref{psidef} and obtain 
\begin{equation} \tilde\Psi(k,t+1)=M_{k}\tilde\Psi(k,t)
\end{equation}
when
\begin{equation}
\sqrt{2}M_{k}=
 \begin{pmatrix}
 -e^{-ik} & e^{-ik} \\
 e^{ik} & e^{ik}
 \end{pmatrix}.
\end{equation}
The eigenvalues of $M_{k}$ are
\begin{align}
 &\lambda^{1}=e^{i\omega_{k}}, &\lambda^{2}=e^{i(\pi-\omega_{k})}
\end{align}
where $\omega_{k}\in[-\pi/2,\pi/2]$ and satisfies $\sin \omega_{k} =\sin
k/\sqrt{2}.$

It follows that
\begin{equation}
\tilde\Psi_{k}(t)=M_{k}^{t}\tilde\Psi(k,0).
\end{equation}

They deduce from this that
\begin{align}
 \tilde\Psi_{\rm{L}}(k,t)&=\frac{1}{2}
  \left(1+\frac{\cos k}{\sqrt{1+\cos^{2}k}}\right)
  e^{i\omega_{k}t} + \frac{(-1)^{t}}{2} 
  \left(1-\frac{\cos k}{\sqrt{1+\cos^{2}k}}\right) e^{-i\omega_{k}t}\\
 \tilde\Psi_{\rm{R}}(k,t)&=\frac{e^{-ik}}{2\sqrt{1+\cos^{2}k}}
  \left(e^{i\omega_{k}}-(-1)^{t}e^{-\omega_{k}t}\right).
\end{align}
Formally inverting the original Fourier transform (using Cauchy's integral
formula) and some ingenious manipulations produce
\begin{align}\label{novel}
 \Psi_{\rm{L}}(n,t)&=\int_{-\pi}^{\pi} \frac{dk}{2\pi}
   \frac{-ie^{ik}}{\sqrt{1+\cos^{2}k}}e^{-i(\omega_{k}t-kn)} \\
 \Psi_{\rm{R}}(n,t)&=\int_{-\pi}^{\pi} \frac{dk}{2\pi}
   \left(1+\frac{\cos k}{\sqrt{1+\cos^{2}k}}\right)e^{-i(\omega_{k}t-kn)}
\end{align}
where $\omega_{k}=\arcsin\left(\frac{\sin k}{\sqrt{2}}\right) \in 
[-\pi/2,\pi/2]$.

They then apply the method of stationary phase to obtain a weaker version 
of Theorem \ref{thmone} above, and integration by parts to show that the 
wavefunction decays superpolynomially fast for $|\alpha|>1/\sqrt{2},$ 
which gives them a much weaker version of Theorem \ref{thmtwo} (Ambainis 
{\emph{et al.}} show that this decay is exponential, but they were unable 
to obtain uniform asymptotics). Of course both approaches consider the 
same functions $\Psi_{\rm{L}}(n,t)$ and $\Psi_{\rm{R}}(n,t)$, the 
differences are just the representation of the generating functions and 
the choice of contour of integration, as we will discuss below. 

If $|\alpha| < \frac{1}{\sqrt{2}}$ each eigenvalue is minus the complex 
conjugate of the other, so the $\psi$-functions have an oscillating 
factor.  When we find the stationary points of the phase, we obtain an 
equation for $k$ at the critical points, $k_{\alpha}.$ This is
\begin{equation}
  \cos k_{\alpha} = \frac{-\alpha}{\sqrt{1-\alpha^2}}.
\end{equation} 
When $|\alpha| < 1/\sqrt{2}$ this has solutions which are real and 
distinct.  The two solutions merge at $\alpha=1/\sqrt{2},$ and then become 
complex.  When $\alpha$ is outside the range $|\alpha| \leq 1/\sqrt{2}$ 
the phase has no stationary point on the real axis.  We have been unable 
to find a method for approximating these integrals.  It is worth noting that 
\eqref{novel} are themselves integral representations of Jacobi polynomials 
as a function of its parameters.  

The exponentially decaying solutions are counter-intuitive in other ways. 
As we mentioned above, for this case $k_{\alpha}$ is complex, so instead 
of seeing the oscillatory behaviour we might be expecting, instead the 
wavefunction decays within an exponential envelope.  This is rather like 
the phenomenon of evanescent waves.  These can also occur classically:  
consider an electromagnetic wave incident on the surface of a conductor.  
These waves cannot propagate in conductors, as the latter will not sustain 
charge gradients.   However, the wave does impinge a finite distance into 
the conductor (the ``skin depth'') over which its amplitude decays 
exponentially.  Mathematically, this is equivalent to a complex wave-number.  
Evanescence can also occur in quantum mechanics, typically in regions that 
are classically forbidden to the particle.  Strangely, both these scenarios 
involve the presence of some kind of barrier, but no such barrier is present 
for the quantum walk. These regions are not classically forbidden to the 
particle, it's just very unlikely to be there.

\subsection{The relationship between the two approaches}

In the paper by McClure and Wong \cite{Wongpaper} the authors
show that the methods of stationary phase can be reduced to the method of 
steepest descent under quite general conditions, as the same results can be
obtained from either method, with exactly the same convergence
properties.  We can see this intuitively as follows.  Using steepest
descents, we have the integral
\begin{equation}\label{steepdef2}
\frac{1}{2\pi i} \int_{\Gamma} g (\zeta) e^{-mh(\zeta)} \; d\zeta.
\end{equation}
If the contour $\Gamma$ goes through the saddlepoint $\zeta^+$ we can 
deform $\Gamma$ to the curve $|\zeta|=|\zeta^+|=r$ or  $\zeta=re^{i\theta},$ 
for some dummy variable $\theta.$ This produces an integral
\begin{equation}
 \frac{1}{2\pi}\int_{-\pi}^{\pi} e^{-mf(re^{{i\theta}})} g(re^{i\theta})
  re^{i\theta} \; d\theta.
\end{equation}
If we separate $f$ into its real and imaginary parts, then we obtain
\begin{equation}
  \frac{1}{2\pi}\int_{-\pi}^{\pi}e^{-m(\Re(f(re^{i\theta}))
  +i\Im(f(re^{i\theta})))}g(re^{i\theta})re^{i\theta}\;d\theta.
\end{equation}
For a quantum system undergoing unitary evolution, $\Re(f)=0$ and since 
$r$ is a constant, this integral is of the form required for the 
stationary phase approximation.  So we will write
$\mu(\theta)=\Im(f(re^{i\theta})).$

The stationary points are those $\theta$ for which $\mu(\theta)'=0$ but 
since
\begin{equation}
  \frac{d\mu(\theta)}{d\theta} = h'(\zeta) \frac{d \zeta}{d \theta}
\end{equation}
we see that the stationary points are identical to the saddle-points. 
With the method of steepest descents the integrand has a very small 
absolute value away from the saddle-point.  By contrast, in the method of 
stationary phase the oscillations of the kernel $\mu$ become arbitrarily 
rapid away from the stationary point, and so self-cancel so long as 
$g(re^{i\theta})$ is sufficiently smooth.  (Readers requiring further 
details are referred to the lucid exposition in \cite{Wongpaper}.)

\section{Conclusion}

We have developed a new way of analysing the discrete quantum walk on the
infinite line in terms of Airy functions, which has the advantage of being
able to handle the dramatic changes in the asymptotic behaviour of this
system in a uniform manner.  We have also probed the mathematical 
relationship between the path-integral and Schr\"{o}dinger approaches to 
solving this problem.  Previous authors have found the methods of 
integration by parts and stationary phase to be problematic over some 
parts of the range of $\alpha.$  By contrast, the method of steepest descents 
yields a unified treatment of the system. 

\bigskip

\subsection*{Acknowledgments}
We are grateful to Rod Wong for drawing our attention to reference 
\cite{Wongpaper} and Nico Temme for kindly generating the steepest 
descent curves.  HAC was supported by MITACS, The Fields Institute, and 
the NSERC CRO project ``Quantum Information and Algorithms.'' The research
of MEHI was partially supported by NSF grant DMS 99-70865.  The research 
of LBR was partially supported by an NSERC operating grant.  
We are also grateful to an anonymous referee for a very thorough reading 
of this paper which helped us to clarify the notation considerably. 



\end{document}